\documentclass[amsmath,
showpacs,prb,twocolumn] {revtex4-1}
\usepackage{bm}
\usepackage{enumerate}
\usepackage{graphicx}
\usepackage[dvips]{epsfig}
\usepackage{epsf}
\makeatletter
\newcommand{\beq}{\begin{equation}}
\newcommand{\eeq}{\end{equation}}
\newcommand{\bea}{\begin{eqnarray}}
\newcommand{\eea}{\end{eqnarray}}
\newcommand{\bse}{\begin{subequations}}
\newcommand{\ese}{\end{subequations}}
\newcommand{\bwt}{\begin{widetext}}
\newcommand{\ewt}{\end{widetext}}
\newcommand{\Rmnum}[1]{\expandafter\@slowromancap\romannumeral #1@}
\newcommand{\nn}{\nonumber}
\makeatletter

\begin{document}

\title{Dynamical spin structure factor of one-dimensional interacting fermions}
\author{Vladimir A. Zyuzin and Dmitrii L. Maslov}
\affiliation{Department of Physics, University of Florida, P.O. Box 118440, Gainesville, Florida 32611-8440, USA}
\begin{abstract}
We revisit the dynamic spin susceptibility, $\chi(q,\omega)$, of one-dimensional interacting fermions.
To second order in the interaction, backscattering results in a logarithmic correction to $\chi(q,\omega)$ at $q\ll k_F$, even if the single-particle spectrum is linearized near the Fermi points. Consequently,  the dynamic spin structure factor, $\mathrm{Im}\chi(q,\omega)$, is non-zero at frequencies above the single-particle continuum.  In the boson language,
this effect results from the marginally irrelevant backscattering operator of the sine-Gordon model. Away from the threshold,
the high-frequency tail
of
$\mathrm{Im}\chi(q,\omega)$ due to backscattering is larger than that due to finite mass by a factor of $k_F/q$. We derive the renormalization group  equations for the coupling constants of the $g$-ology model at finite $\omega$ and $q$ and find the corresponding expression for  $\chi(q,\omega)$, valid to all orders in the interaction but not in the immediate vicinity of the continuum boundary,
where the finite-mass effects become dominant.
\end{abstract}
\maketitle

\noindent
{\it \underline{Introduction}}
Bosonization is the most common way to
describe
one-dimensional (1D) interacting fermions.\cite{Giamarchi}
If the lattice effects are not essential, an exact correspondence between the fermion
and fermion-hole (boson)
operators
maps the charge sector of the system onto a gas of free bosons [the Tomonaga-Luttinger liquid (TLL)].
The spin sector, however, is not free but maps onto the sine-Gordon model.
 The non-Gaussian (cosine) term of this model results from backscattering of fermions with opposite spins.
If the interaction between the original fermions is repulsive, the backscattering term represents a marginally irrelevant operator
 and is renormalized down to zero at the fixed point, where the spin sector also becomes free.
At intermediate energy scales,
such marginally irrelevant operators
lead to logarithmic renormalizations of
the
observables. \cite{Cardy_86} Since
 the original paper by Dzyaloshinskii and Larkin (DL),~\cite{DL72} it has been known that
the backscattering operator gives rise
 to the logarithmic temperature (or external magnetic field) corrections to the static spin susceptibility. In Refs.~\onlinecite{BKV_prb97,CM_prb03}, it was shown that the static spin susceptibility also depends logarithmically on the external momentum $q$
at small $q$. In addition, both the spin- and charge
 susceptibilities at $2k_F$ acquire multiplicative logarithmic renormalizations.  \cite{GS_prb89,Giamarchi}

\begin{figure}[t]
\includegraphics[width=0.8 \linewidth]{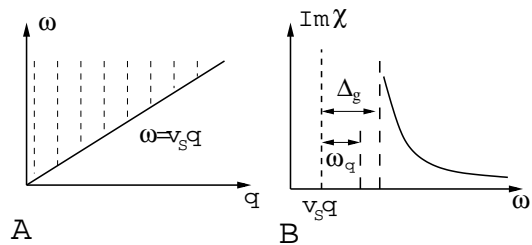}
\caption{A. Particle-hole excitations at small momenta and frequency in 1D fermion system. The line $\omega =v_{F}q$ corresponds to a continuum. The hatched area is the domain of incoherent spin fluctuations  arising due to backscattering processes. B.
Schematic frequency dependence of the DSSF at a given momentum $q$.
Renormalization group
flows into the strong coupling regime at
frequency $\Delta_{g}$
[Eq.~(\ref{Deltag})]
above the threshold.
At distance $\omega_{q} = q^2/m$ to the threshold, the effect of finite mass becomes more important than
that of
backscattering. \label{fig1}}
\end{figure}

In this work, we focus on  {\it dynamics} of the long-wavelength part of the spin response.
First, we need to outline the differences between the charge and spin sectors.
As charge bosons are free at all energies, the dynamical charge structure factor
(the imaginary part of the charge susceptibility at finite frequency, $\omega$, and momentum, $q$)
is a delta function centered at the boson dispersion, $\omega=v_c q$, which is represented by a straight line in Fig.~\ref{fig1}A. This result differs from that for
free fermions only in that  the Fermi velocity, $v_F$, is replaced by the renormalized charge velocity, $v_c$. A non-zero width of the charge structure factor appears only if one goes beyond the TLL paradigm by taking into account finite curvature (inverse mass) of the fermion spectrum.
In the pioneering paper~\cite{Pustilnik_prl06} and subsequent work
(for a review, see Ref.~\onlinecite{ISG}), it was shown that the combined effect of the curvature and  interactions results
in many new features in the charge structure factor,
such as
edge singularities at the boundaries of the continuum and the high-frequency tail both of which are absent for free fermions.

As far as the spin channel is concerned,  it is
common to treat
the backscattering operator via renormalization group (RG).
As the fixed point corresponds to free bosons, the formal result for the dynamical spin structure factor (DSSF) at the fixed point
 is also a delta function with $v_c$ replaced by the spin velocity, $v_s$.
 The problem with this argument is that DSSF is measured at finite $\omega$ and $q$ and thus away from the fixed point. Therefore, its dependences on $\omega$ and $q$ must reflect the flow at finite rather than infinite RG time.

In this paper, we revisit the
DSSF of a 1D interacting fermion system. Besides being of a fundamental interest on its own, the DSSF is relevant for a number of experiments in both condensed-matter and cold-atom systems, such  inelastic neutron and spin-resolved X-ray spectroscopies, nuclear magnetic resonance, spin Coulomb drag, \cite{spindrag} etc.   First, we show by direct perturbation theory that the logarithmic renormalization of the dynamical spin susceptibility occurs in a Lorentz-invariant way
via a  $\ln(v_F^2q^2-\omega^2)$ term. This already implies that, in contrast to the free-fermion case, the DSSF is non-zero in the entire sector $|\omega|>v_F|q|$ (the hatched region in Fig.~\ref{fig1}A). In contrast to the charge sector, this high-frequency tail occurs even without taking into account the finite-curvature effects. Next, we collect all leading logarithmic terms by using RG for the spin vertex. Our final result for the spin susceptibility reads
\begin{eqnarray}
\label{result0}
\chi(q,\omega)&=&\chi_0 \frac{v_sv_{F}q^2}{(v_sq)^2-\omega^2-i\delta}\label{result}\\
&\times&
\left\{1+
\frac{g_{1}}{2}  \frac{1}{1+\frac{g_{1}}{2}\ln\left[\frac{(v_{s}\Lambda)^2}{(v_{s}q)^2-\omega^2-i\delta}\right]}\right\},\nonumber
\end{eqnarray}
where
$\chi_0=
2/\pi v_F$ is the spin susceptibility of 1D free fermions, $g_1$ is the (dimensionless) backscattering amplitude,  $\Lambda$ is the ultraviolet cutoff,
$\delta= \mathrm{sgn}\omega 0^+$, and \begin{equation}
\label{vs}
v_s=v_F\sqrt{1-g_1}
\end{equation}
 is the spin velocity. \cite{CMS_prb08,comment_giamarchi} At $\omega=0$ and to second order in $g_1$, Eq.~(\ref{result}) reduces to the $\ln q$ correction of Refs.~\onlinecite{BKV_prb97,CM_prb03}. Also, setting $\omega=0$ and replacing $v_Fq$ by either temperature or the Zeeman energy in the external magnetic field, we reproduce the DL result.\cite{DL72} To second order in $g_1$, our result
is in line with those of
Refs.~\onlinecite{Eggert_prl94,Pereira_prl06} for a spin-$1/2$ Heisenberg antiferromagnetic chain.
In that case, the high-frequency tail of the DSSF occurs due to a marginally irrelevant operators arising from
umklapp
scattering.
The profile of $\mathrm{Im}\chi(q,\omega)$ is shown schematically in Fig.~\ref{fig1}B. Sufficiently close to the threshold,
 the divergence in  $\mathrm{Im}\chi(q,\omega)$ must be regularized by the finite-curvature effects--see below for a more detailed discussion of the crossover between the high-frequency tail and threshold singularities.

\begin{figure}[t]
\includegraphics[width=0.8 \linewidth]{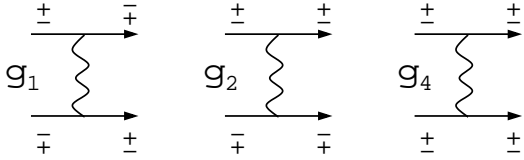}
\caption{Scattering amplitudes of the fermion-fermion interaction $g_{1,2,4}$. \label{fig2}}
\end{figure}

\noindent
{\it \underline{Feynman diagrams in the fermion language.}}
As usual, we linearize the fermion spectrum near the Fermi energy and decompose the fermion operators into the left- and right-movers, so that the Hamiltonian of a free system is
\begin{equation}
H =\sum_{s=\pm,\alpha} \int
dp~ \xi_{s}(p)
c^{\dag}_{s,\alpha}(p)
c_{s,\alpha}(p),
\end{equation}
where $\xi_{\pm}(p) = \pm v_{F}p$, with $p$ being a deviation from the Fermi momenta, $v_{F}$ the Fermi velocity,
and $c_{\pm,\alpha}(p)$
the right/left-moving fermion
with spin $\alpha$.
Linearization presumes
that the fermion momenta
are bounded by $\pm \Lambda_{F}$.
Fermions interact via a short-range and SU$(2)$-invariant potential $U(q)$,
parameterized
by three (dimensionless)
scattering amplitudes, $g_{1}$, $g_{2}$, and $g_{4} $, which are
defined in Fig.~\ref{fig2}.
To first order in $U$,
$g_2=g_4=U(0)/\pi v_F$ and
$g_1=U(2k_F)/\pi v_F$.
We assume that the Fermi momentum is not commensurate with the lattice and thus neglect umklapp scattering.

\begin{figure}[htp]
\vspace{0.1in}
\includegraphics[width=0.8 \linewidth]{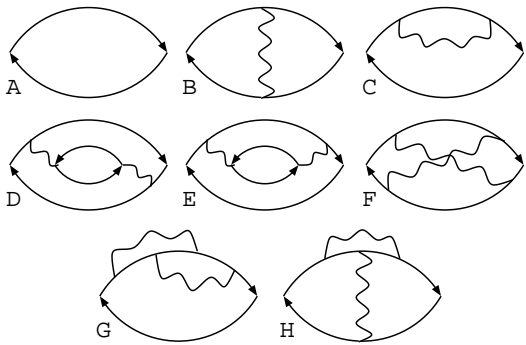}
\caption{Feynman diagrams
for the spin susceptibility
to second order in the interaction (wavy line). The Aslamazov-Larkin diagrams (not shown) vanish identically by SU$(2)$ symmetry. The second-order RPA diagram (also not shown), obtained from diagram B by inserting one more wavy line parallel to the first one, does not contain a logarithmic singularity and is thus ignored.
\label{fig3}}
\end{figure}

We now calculate the spin susceptibility at small but finite $\omega$ and $q$ via a perturbation theory in the coupling constants $g_1$, $g_2$, and $g_4$.
The
free spin susceptibility--diagram A in Fig.~\ref{fig3}-- is given by
\begin{align}
\label{chi0}
\chi^{(0)}(q,\omega) = \chi_{0} \frac{(v_{F}q)^2}{\omega^2 + (v_{F}q)^2},
\end{align}
where $\omega$ is a Matsubara frequency.
Upon analytic continuation to real frequencies ($\omega\to-i\omega+0^+$),
the imaginary part of the susceptibility is $\propto q^2 \delta(\omega^2 - v^2_{F}q^2)$, which corresponds to well-defined spin excitations.

The first-order corrections are given by diagrams \ref{fig3}B and \ref{fig3}C.
It can be shown [see Supplementary Material (SM)]
that diagrams with $g_{4}$ sum up to zero, while diagrams with $g_{2}$ cannot be constructed at this order. For the backscattering contribution, we obtain
\begin{align}\label{result1}
\delta \chi^{(1)}(q,\omega) =  g_{1}\chi_{0}\bigg[\frac{(v_{F}q)^2}{\omega^2 + (v_{F}q)^2}\bigg]^2.
\end{align}

To
second order in the interaction, there are five non-trivial diagrams for the spin susceptibility, presented in Fig.~\ref{fig3}D-H. Calculations show that all the $g_{1}g_{2}$  terms
from diagrams \ref{fig3}F-H
sum up to zero. All the $g_{4}^2$ terms
from all the second-order diagrams
sum up to zero as well. Finally,
the $g_{2}^2$ terms
from
diagrams \ref{fig3}D and \ref{fig3}E cancel each other.
What remains
is the backscattering, $g_1^2$ contribution
from diagrams \ref{fig3}D and \ref{fig3}E. In the leading logarithmic approximation, we find
\begin{align}\label{result2}
\delta \chi^{(2)}(q,\omega) = -\frac{1}{4}g_{1}^{2}\chi_{0} \frac{(v_{F}q)^2}{\omega^2 + (v_{F}q)^2}  \ln\bigg[ \frac{(v_{F}\Lambda)^2}{\omega^2 + (v_{F}q)^2}  \bigg],
\end{align}
where $\Lambda$ is the cutoff imposed on the interaction.
At $\omega=0$, Eq.~(\ref{result2}) reduces to the result of
Refs.~\onlinecite{BKV_prb97,CM_prb03}.
Upon analytic continuation,
the logarithmic factor gives rise to a non-zero DSSF at $|\omega|>v_F|q|$.
In the next section, we will employ RG
to calculate the spin susceptibility to first order in the renormalized
vertex. We outline the resulting expression for the spin susceptibility:
\begin{eqnarray}
\label{dm1}
\chi(q,\omega)
=\chi_0\left\{\frac{(v_Fq)^2(1-g_1/2)}{\omega^2+(v_Fq)^2}
+g_1\frac{(v_Fq)^4}{\left[\omega^2+(v_Fq)^2\right]^2}
\right. \nonumber \\ \left.
+\frac{g_{1}}{2}\frac{(v_{F}q)^2}{\omega^2 + (v_{F}q)^2} \frac{1}{1+\frac{g_{1}}{2}\ln\left[\frac{(v_{F}\Lambda)^2}{\omega^2 + (v_{F}q)^2}\right]}
\right\},
\end{eqnarray}
where a factor of $-g_1/2$ in the numerator of the first term was added to compensate for the first-order contribution from the third term. The first and second terms in Eq.~(\ref{dm1}) can be combined into an expansion of
$\chi^*(q,\omega)=\chi_0v_sv_Fq^2/\left[\omega^2+(v_sq)^2\right]$ to order $g_1$. Note that $\chi^*(q,\omega)$ is the spin susceptibility of free bosons,
$\chi_b(q,\omega)=K_s \chi^*(q,\omega)$,  at the fixed point, where the Luttinger parameter of the spin channel, $K_s$, is renormalized to unity. We surmise that all non-logarithmic terms can be absorbed into $\chi^*(q,\omega)$. As far as the last term in Eq.~(\ref{dm1}) is concerned, we cannot distinguish between $v_s$ and $v_F$ within the leading logarithmic approximation. However, guided by the general RG principle,
we conjecture that $v_F$ in the last term must be replaced by $v_s$ as well. Performing these replacements, we obtain (after analytic continuation) the result announced in Eq.~(\ref{result}).
Next, we are going to show that RG does indeed reproduce the logarithmic part of the result in Eq.~(\ref{dm1}).

\begin{figure}[t]
\includegraphics[width=0.8 \linewidth]{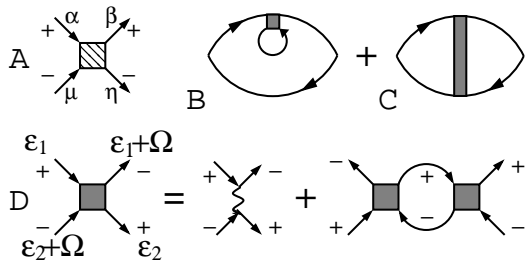}
\caption{A.
Interaction vertex
given by Eq.~(\ref{interaction_vertex}). B,C. Spin susceptibility to first order in
renormalized
vertex.
D. Graphical equation for the backscattering
amplitude $\gamma_{1}$.
\label{fig4}}
\end{figure}

\noindent
{\it \underline{Renormalization group.}}
From now on, we neglect the $g_{4}$ processes,
as they do not flow
under RG.
The interaction vertex,
shown graphically in Fig.~\ref{fig4}A, can be decomposed into the spin and charge parts as
\begin{align} \label{interaction_vertex}
\frac{1}{\pi v_F}
\Gamma^{\alpha\beta}_{\mu\eta} = -\frac{1}{2}\gamma_{1} {\vec \sigma}_{\alpha\beta} \cdot{\vec \sigma}_{\mu\eta} - \left(\frac{1}{2}\gamma_{1}  - \gamma_{2}\right) \delta_{\alpha\beta} \delta_{\mu\eta},
\end{align}
where
$\gamma_{1,2}$ are the renormalized back- and forward-scattering amplitudes.
As these vertices will be used to find the spin susceptibility at finite $\omega$ and $q$, we will need to know them away from the Fermi surface along both the energy and momentum axes. The equations for $\gamma_{1}$ and $\gamma_2$, derived in SM
and shown graphically in Fig.~\ref{fig4}D, are of the standard form\cite{solyom,Giamarchi,M_review}
\begin{eqnarray}
\label{rg}
\frac{d\gamma_{1}(\ell)}{d\ell}  = -\gamma_{1}^2(\ell),\quad
\frac{d\gamma_{2}(\tilde\ell)}{d\tilde \ell}  = -\frac{\gamma_{1}^2(\tilde\ell)}{2},
\end{eqnarray}
except that the RG times, $\ell$ and $\tilde\ell$, are different.
The vertex $\gamma_1$, which is renormalized in the particle-hole channel, evolves with $\ell \equiv \ln\left[ \frac{
v_{F}\Lambda
}{\sqrt{\Omega^2 + (v_{F}Q)^2}}
\right]$, where $\Omega$ and $Q$ are the energy and momentum transfers through the vertex, correspondingly.
The vertex $\gamma_2$, which is renormalized in the particle-particle channel, depends on the total incoming energy and momenta. The meaning of Eq.~(\ref{rg}) is that one first solves for $\gamma_1$ as a function of $\ell$ and then replaces $\ell$ by $\tilde\ell$ to find $\gamma_2$. The
initial values are $\gamma_{i}(\ell =0) = g_{i}$, $i=1,2$.  (Details of the derivation are given in SM.)
Solving Eq.~(\ref{rg}), one obtains
\begin{align}\label{gamma1}
\gamma_{1}(\ell) = \frac{g_{1}}{1+g_{1}\ell},~~ \gamma_{1}(\tilde\ell)-2\gamma_{2}(\tilde\ell) = \mathrm{const}.
\end{align}
We are now in a position to calculate
the renormalized
 spin susceptibility at finite $\omega$ and $q$, using
$\gamma_{1}$ as an effective interaction.
To first order in $\gamma_{1}$, there are only two diagrams
 for the spin susceptibility: diagrams B and C in Fig.~\ref{fig4}.
 Combining the two diagrams
 together and using Eq.~(\ref{gamma1}),
 we obtain
 renormalized the part of the spin susceptibility
 along with
 the linear in $g_{1}$ term (see SM for details of the calculations):
\begin{align}\label{result6}
\chi^{(\ln)} &= -4\bigg[\frac{(v_{F}q)^2}{\vert{\bar\omega}\vert^4}\bigg]\int\frac{dQd\Omega}{(2\pi)^2}
\ln\bigg[\frac{(v_{F}\Lambda)^2}{\vert {\bar \Omega} + {\bar \omega}\vert^2}\bigg]\gamma_{1}(\ell)
\nonumber
\\
&
= \chi_{0}\frac{g_{1}}{2} \frac{(v_{F}q)^2}{\omega^2+(v_{F}q)^2} \frac{1}{1+\frac{g_{1}}{2}\ln\bigg[\frac{(v_{F}\Lambda)^2}{\omega^2+(v_{F}q)^2}\bigg]},
\end{align}
where ${{\bar \Omega}} = \Omega+iv_{F}Q$ and ${\bar \omega} = \omega+iv_{F}q$.
This result indeed coincides with the last term in
Eq.~(\ref{dm1}), and thus the conjecture leading to Eq.~(\ref{result}) is proven.

Equation (\ref{result}) is the
central result of
this paper, which shows that DSSF is non-zero at $|\omega|>v_F|q|$. Away from the free-boson pole,
the DSSF is given by
\begin{eqnarray}\label{result5}
&\mathrm{Im}
\chi(q,\omega)=
 \chi_0\frac{\pi g_{1}^2}{4}\frac{v_Fv_sq^2}{\omega^2-(v_sq)^2}
 \frac{\mathrm{sgn}\omega}
 {\left[1+\frac{g_{1}}{2}\ln\left[\frac{(v_{
 s}\Lambda)^2}{\omega^2 - (v_{
 s}q)^2}\right] \right]^2 + \left[\frac{\pi g_{1}}{2}\right]^2}.
\nonumber\\
\end{eqnarray}
The conditions for the validity of Eq.~(\ref{result5}) near the threshold are discussed before the concluding paragraph of this paper.

\noindent
{\it \underline{Fermi liquid in 1D.}}
Putting $\omega=0$, replacing
$v_Fq\to \max\{T,H\}$ under the logarithm, and expanding back the non-logarithmic term to order $g_1$ in Eq.~(\ref{result}), we reproduce  the DL result  for the
static spin susceptibility:\cite{DL72}
\begin{align}\label{DLresult}
\chi(T,H)= \chi_{0}\left\{1 + \frac{g_{1}}{2}\left[ 1+\frac{1}{1+g_{1}\ln\left[\frac{v_{F}\Lambda}{\mathrm{max}(T,H)}\right]} \right]\right\}.
\end{align}
We now show that, somewhat counterintuitively, the DL result can be reproduced by the Fermi-liquid (FL) theory, if one replaces the FL parameters by running values of the scattering amplitudes.
We recall the FL expression for the spin susceptibility \cite{LL9}
\begin{equation}\label{FermiSusceptibility}
\chi = \chi_{0}\frac{ m^{*}}{m}\frac{1}{1+\Gamma_s\frac{m^{*}}{m}Z^2},
\end{equation}
where $m^{*}=m\left[1-\partial_\omega\Sigma(\omega,k)\right]/\left[1+\partial_k\Sigma(\omega,k)/v_F\right]\vert_{\omega,k=0}$ is the effective mass, $Z=\left[1-\partial_\omega\Sigma(\omega,k)\right]^{-1}\vert_{\omega,k=0}$ is the quasiparticle residue,
and $\Gamma_s$ is the spin part of the interaction vertex.
 Since the original interaction is static, it gives a frequency-independent self-energy to first order. Therefore, the $Z$-factor is not renormalized to this order, while the momentum dependence of the self-energy amounts to renormalization of the effective mass
$m^{*} = m(1+g_{1}/2)$ both for right- and left-moving fermions. According to Eq. (\ref{interaction_vertex}), $\Gamma_s=-\gamma_1/2$,
where $\gamma_1$ is given by  Eq.~(\ref{gamma1}) with $\ell = \ln\left[\frac{v_{F}\Lambda}
{\mathrm{max}(T,H)}\right]$ in the static case.
Substituting these $m^*$ and $\Gamma_s$ into Eq.~(\ref{FermiSusceptibility}) and expanding to first order in $\gamma_1$, we reproduce the DL result, Eq.~\ref{DLresult}.  Although the FL theory is not valid in $D=1$, it is still valid in $D=1+\varepsilon$ with $\varepsilon\to 0^+$. Our example shows that a logarithmic flow of the scattering amplitude with $\varepsilon$ reproduces the correct result even in $D=1$.

Another interesting feature of the DL result is that the fixed-point value $\chi=\chi_0(1+g_1/2)$ is renormalized by $g_1$. This seems to contradict RG because $g_1$ flows to zero at the fixed point. In fact, this means that not all the coupling constants in the perturbative result should be replaced by the running values: some remain at their bare values evaluated at the ultraviolet cutoff. This is an example of the ``anomaly'' frequently encountered in massless field-theoretical models. \cite{gross} Another example of such an anomaly is the specific heat of 1D fermions.~\cite{CMS_prb08}

\noindent
{\it \underline{Connection to bosonization.}}
Certainly, the fermion language is not a preferred one: one can equally well obtain the same results in the boson language.
However, one has to be careful about not taking the limit of local interaction too soon. Assuming a non-local interaction potential $U(x-x')$, we bosonize the Hamiltonian, treating
the interaction as local in the Gaussian part but keeping it non-local in the non-Gaussian part. For clarity, we distinguish between the backscattering amplitudes of fermions with the same and opposite spins, $g_{1||}$ and $g_{1\perp}$, correspondingly. This yields \cite{CMS_prb08}
\begin{eqnarray}
&H = \frac{1}{2\pi} \sum_{i=c,s}\int dx
 \bigg[   \frac{
 v_{i}}{K_{i}}(\partial_{x}\phi_{i})^2 +
 v_{i}K_{i}(\partial_{x}\theta_{i})^2  \bigg]
 \nonumber\\
&+\frac{
\Lambda_F^2}{
2\pi^2
}\sum_{\alpha=\pm} \int \int dxdx^\prime ~ U(x-x^{\prime})
\nonumber\\
& \times \cos\left\{\sqrt{2}\left[\phi_{s}(x)
+\alpha \phi_{s}(x^{\prime})\right]\right\}
 \nonumber\\
&\times \cos\left\{\sqrt{2}\left[\phi_{c}(x)-\phi_{c}(x^{\prime})\right] +2k_{F}(x-x^{\prime})\right\},
 \label{backscattering}
\end{eqnarray}
where $\phi_{c(s)}$ and $\theta_{c(s)}$ are usual position and momentum boson fields in the charge ($c$) and spin ($s$) sectors, $v_s=v_F$, $K_s=1$, and explicit expressions for $v_c$ and $K_c$ are given in SM.
The standard form of the Hamiltonian is reproduced if the local limit is taken also in the non-Gaussian part, upon which we obtain a spin-charge separated sine-Gordon model with the coupling $g_{1\perp}$ in the cosine term and renormalized spin velocity $v_{s} = v_{F}\sqrt{1-g_{1\parallel}}$ and Luttinger parameter $K_{s} = 1/\sqrt{1-g_{1\parallel}}$. With Hamiltonian (\ref{backscattering}), one can construct a perturbation theory for the spin susceptibility $\chi(x,\tau)=-\langle T_{\tau}\partial_x\phi_s(x,\tau)\partial_x\phi_s(0,0)\rangle$. In doing so, we reproduce the same diagrams for the spin susceptibility as in the fermion approach.
The first-order term of the fermion approach, Eq.~(\ref{result1}), is reproduced correctly only if one keeps non-local interaction in the non-Gaussian term. The reason is that a part of this result comes from mass renormalization (diagram C in Fig.~\ref{fig3}), which is absent to first order in the local interaction. Starting from second order, one can take the local limit. The results of the boson and fermion approaches are identical, as they should be. In particular, the second-order result for $\chi(q,\omega)$ [Eq.~(\ref{result2})] can obtained by expanding the partition function of the sine-Gordon model to second order in the backscattering operator, as it was done in Ref.~\onlinecite{Pereira_prl06} for the umklapp operator. In our case,  this gives $\chi(q,\omega)\propto q^2(\omega^2+v^2_sq^2)^{2K_s-3}$ which, upon expanding near weak coupling ($K_s\approx 1+g_{1||}/2$), reproduces Eq.~(\ref{result2}).

\noindent
{\it \underline{
Finite mass effects.}}
In the charge sector, taking into account finite curvature of the fermion spectrum is the only way to smear the delta-function singularity
of the dynamic charge structure factor.\cite{Pustilnik_prl06,ISG} In the spin sector,  there are  two competing effects that lead to a non-zero DSSF outside the continuum: the marginally irrelevant backscattering operator and finite curvature.
Near the the threshold $|\omega|=v_s|q|$,  the spin susceptibility in Eq.~(\ref{result0}) diverges and thus the perturbation theory breaks down. The divergence must be  regularized by finite curvature of the single-particle spectrum. For massive fermions with dispersion $p^2/2m$, the effect of finite curvature becomes important when the distance to the threshold $\Delta\equiv |\omega|-v_s|q|$ becomes comparable to $\omega_q\equiv q^2/m$. On the other hand, the RG flow for massless fermions becomes important at $\Delta$ becomes comparable to
\begin{align}
\Delta_g\equiv (v_s\Lambda^2/q)\exp(-2/g_1).\label{Deltag}
\end{align}
RG for fermions is thus valid if $\Delta_g\gg \omega_q$, i.e., if $q\gg (mv_s\Lambda^2)^{1/3}\exp(-2/3 g_1)$.  In the opposite case ($\Delta_g\ll \omega_q$), the finite-mass effects become relevant before the RG flow develops. See Fig. \ref{fig1} for schematics.

Even in the case of $\Delta_g\gg \omega_q$, matching threshold singularities due to finite mass for spinful fermions (Ref.~\onlinecite{schmidt:2010}) with the RG result of this paper still remains an open question. However, one can compare the high-frequency tails due to each of these effects. For $\Delta\gg \Delta_g$, the RG result reduces back to the second-order one. Up to a numerical coefficient, which is irrelevant for the present discussion,
\beq
\label{HFRG}
\mathrm{Im}\chi(q,\omega)\sim \chi_0 g_1^2 \frac{v^2_Fq^2}{\omega^2-v_F^2q^2}\Theta(|\omega|-v_F|q|).
\eeq
On the other hand, the high-frequency tail arising from finite mass
can be calculated via perturbation theory in $1/m$ (valid for $\Delta\gg \omega_q$).  Such a calculation was carried out in Refs.~\onlinecite{Pustilnik_prl06,Pereira_prl06,Teber} for the charge susceptibility of spinless fermions. For massive fermions, however, the difference in the $1/m$ expansions of the charge and spin susceptibilities amounts only to a numerical coefficient. Up to this coefficient, we can simply borrow the result from Refs.~\onlinecite{Pustilnik_prl03,Pereira_prl06,Teber}, which reads
\beq
\label{HFM}
\mathrm{Im\chi}_{1/m}(q,\omega)\sim g^2 \left(\frac{q}{mv_F}\right)^2 \frac{v^2_Fq^2}{\omega^2-v_F^2q^2}\Theta(|\omega|-v_F|q|),
\eeq
where $g$ is some dimensionless coupling constant. Comparing Eqs.~(\ref{HFRG}) and (\ref{HFM}), we see that the high-frequency tail due to a marginally irrelevant operator is larger than that due to finite mass by factor of $mv_F/q\gg 1$.

\noindent
{\it \underline{Conclusions.}}
To conclude, we have studied the dynamical spin structure factor (DSSF) of one dimensional interacting fermions
for small momenta ($q\ll k_F$). In contrast to the charge structure factor, the DSSF is non-zero above the continuum even in a model with linearized fermion spectrum due to the effect of a marginally irrelevant backscattering operator. We found the DSSF by direct perturbation theory in the fermion language, supplemented by RG. The high-frequency tail due to backscattering is larger than that due to finite mass by a factor of $mv_F/q\gg 1$. One immediate application of our result is the non-analytic temperature dependence of the nuclear spin relaxation rate $T_{1}^{-1}\propto T \int dq \frac{\mathrm{Im}\chi(q,\omega)}{\omega}\Big\vert_{\omega\to 0}$
resulting from the region $q\ll k_F$. Logarithmic renormalization of $\chi(q,\omega)$ modifies the Korringa law as $1/T_{1} \propto T/\ln(v_{F}\Lambda/T)$. At weak coupling, this renormalization makes the $q=0$ contribution comparable to the usually considered $2k_F$ contribution.\cite{Giamarchi}

We are grateful to A. V. Chubukov, S. Maiti, C. Reeg, O. A. Starykh, and especially to L. I. Glazman for stimulating discussions. This work was supported by the National Science Foundation via grant NSF DMR-1308972.

\newpage
\begin{widetext}
\appendix
\begin{center}
{{\bf Supplementary material}}
\end{center}

\section{Diagrams for the spin susceptibility of 1D fermions}
\setcounter{page}{1}
In the section, we present the calculation of the dynamic spin susceptibility to second order in the interaction. We consider scattering processes shown in Fig.~\ref{fig2} of the the main text (MT) and neglect Umklapp scattering. For an $SU(2)$-invariant interaction $U(q)$, $g_1=U(2k_F)/\pi v_F$ and $g_2=g_4=U(0)/\pi v_F$. However, we will formally treat $g_2$ and $g_4$ as two independent parameters. For convenience, we reproduce here Fig.~\ref{fig3} of MT.

\subsection{Non-interacting fermions}
The Green function of non-interacting 1D fermions is given by
\begin{align}
G_{\pm}(p,\epsilon) = \frac{1}{i\epsilon - \xi_{\pm}(p)},
\end{align}
where $\epsilon$ is the Matsubara frequency, $\pm$ denotes the right/left mover, and $\xi_{\pm}(p) = \pm v_{F}p$. Note that
$G_{-}=-G^*_+$.  The spin susceptibility of non-interacting 1D fermions (diagram A in Fig.~\ref{fig3SM}) can be calculated as
\begin{align}
\chi^{(0)}(q,\omega)=-g_{e}^2\frac{1}{2}\bigg<G_{+}(p,\epsilon)G_{+}(p+q,\epsilon+\omega)\bigg>_{\epsilon,p} + \mathrm{c.c.} =
-\frac{g_e^2}{4\pi i}
\bigg[ \frac{q}{\omega + iv_{F}q} - \frac{q}{\omega - iv_{F}q} \bigg] = \chi_{0}\frac{(v_{F}q)^2}{\omega^2 + (v_{F}q)^2},
\end{align}
where $\chi_{0} =
{g_{e}^2}/{2v_{F}\pi}$ is the static spin susceptibility, $g_{e}\approx 2$
is the $g$-factor,
and where we introduced the following notation
\begin{align}
\bigg<...\bigg>_{\epsilon,p} = \int\frac{d\epsilon dp}{(2\pi)^2}\bigg[...\bigg].
\end{align}
We will always be integrating over $\epsilon$ first and then over
$p$.
An important building block
for all diagrams is
the polarization bubble of left/right-moving fermions:
\begin{align}
\label{pipm}
\Pi_{\pm}(q,\omega)=\bigg<G_{\pm}(p,\epsilon)G_{\pm}(p+q,\epsilon+\omega)\bigg>_{\epsilon,p} = \pm\frac{1}{2\pi i} \frac{ q}{\omega \pm iv_{F}q}.
\end{align}

\begin{figure}[t]
\includegraphics[width=0.8 \linewidth]{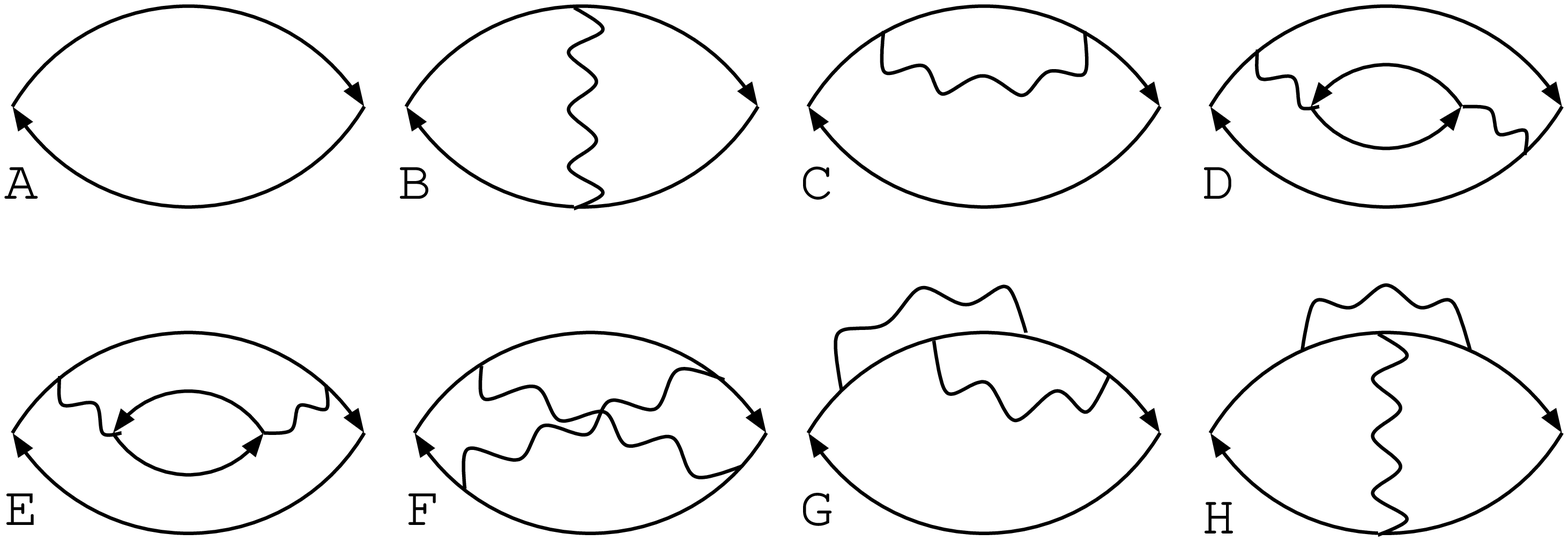}
\caption{Feynman diagrams
for the spin susceptibility
up to second order in the interaction (wavy line).
\label{fig3SM}}
\end{figure}

\subsection{First order}
To first order in the interaction,
there are two diagrams: diagrams B and C in Fig.~\ref{fig3SM}. The interaction in both diagram can be either in the $g_1$ or $g_4$ channel; there are no $g_2$ diagrams to this order.
In the $g_1$ channel, diagrams  B and C give, correspondingly
\bea
\chi^{(1)}_{B,g_1}(q,\omega)&=& \frac{1}{2}g_{1}\chi_{0} \frac{(v_{F}q)^2}{\omega^2 + (v_{F}q)^2}, \nn\\
\chi^{(1)}_{C,g_1}(q,\omega)&=&-\frac{1}{4}g_{1}\chi_{0}  \bigg[ \frac{(v_{F}q)^2}{{\bar\omega}^2} + \frac{(v_{F}q)^2}{({\bar\omega}^{*})^2}\bigg],
\eea
where
\beq\bar\omega=\omega+iv_Fq.
\label{baro}
\eeq
Summing up the two contributions, we obtain
\begin{align}
\label{BC1}
\chi^{(1)}_{B,g_1}(q,\omega)+\chi^{(1)}_{C,g_1}(q,\omega)=g_{1}\chi_{0}\bigg[\frac{(v_{F}q)^2}{\omega^2 + (v_{F}q)^2} \bigg]^2.
\end{align}

In the $g_{4}$ channel, the contributions of diagrams B and C are
\bea
\chi^{(1)}_{B,g_4}&=&-\frac{1}{4}g_{4}\chi_{0}\bigg[ \frac{(v_{F}q)^2}{{\bar \omega}^2} + \frac{(v_{F}q)^2}{({\bar\omega}^{*})^2}\bigg],\nn\\
\chi^{(1)}_{C,g_4}&=&\frac{1}{4}g_{4}\chi_{0}\bigg[ \frac{(v_{F}q)^2}{{\bar \omega}^2} + \frac{(v_{F}q)^2}{({\bar\omega}^{*})^2}\bigg].
\eea
We see
that the two diagrams
cancel each other. The final first-order result is then given by Eq.~(\ref{BC1}) which coincides with Eq.~(\ref{result1}) of MT.

Note that
the difference
in signs of the self-energies in the $g_{1}$ and $g_{4}$ channels can be traced down to the relation
\begin{align}
\int_{-\infty}^{+\infty}\frac{1}{i\epsilon - \xi_{\pm}(p) }\frac{d\epsilon}{2\pi} = -\frac{1}{2} \mathrm{sign}(\xi_{\pm}(p)) = \mp\frac{1}{2}\mathrm{sign}(p).
\end{align}

\subsection{Second order}
In the $g_1^2$ channel, the second order ladder-type diagram
shown
in Fig.~\ref{fig3SM}D yields
\begin{align}
\chi^{(D)}_{1}(q,\omega) = 8U^{2}(2k_{F}) \bigg<{\cal T}^{(D)}_{+}(Q,\Omega, q,\omega){\cal T}^{(D)}_{-}(Q,\Omega, q,\omega)  \bigg>_{\Omega, Q},
\end{align}
where a factor of two accounts for contributions from left- and right-moving internal fermions,
\begin{align}
{\cal T}_{+}^{(D)}(Q,\Omega,q,\omega) = \bigg<G_{+}(p,\epsilon)G_{+}(p+Q,\epsilon+\Omega)G_{+}(p+q,\epsilon+\omega)\bigg>_{\epsilon,p} = -\frac{i}{{\bar \omega}}\frac{1}{2\pi}\bigg[\frac{Q}{{\bar \Omega}} - \frac{Q-q}{{\bar \Omega} - {\bar \omega}}\bigg],
\end{align}
and
\bea
{\cal T}_{-}^{(D)}(Q,\Omega,q,\omega) &=& \bigg<G_{-}(k,\epsilon^\prime)G_{-}(k+Q,\epsilon^\prime+\Omega)G_{-}(k+q,\epsilon^\prime+\omega)\bigg>_{\epsilon^\prime,k} =-\left[{\cal T}_{+}^{(D)}(Q,\Omega,q,\omega)\right]^*\nn\\
&&= -\frac{i}{{\bar \omega}^{*}}\frac{1}{2\pi}\bigg[\frac{Q}{{\bar \Omega}^{*}} - \frac{Q-q}{{\bar \Omega}^{*} - {\bar \omega}^{*}}\bigg]
\eea
with
\beq
{\bar \Omega} = \Omega + iv_{F}Q.
\label{barO}
\eeq
Starting from the second order, diagrams for the spin susceptibility contain a logarithmic divergence which is cut off by external $q$ and $\omega$. When opening the product ${\cal T}^{(D)}_+{\cal T}^{(D)}_-$, we keep only the logarithmically divergent terms. This yields
\begin{align}
\chi^{(D)}_{1}(q,\omega) = -U^{2}(2k_{F}) \frac{8}{\vert{\bar\omega}\vert^2} \frac{1}{(2\pi)^2}\bigg< \frac{Q(Q-q)}{{\bar \Omega}({\bar \Omega}^{*}-{\bar \omega}^{*})}  +
\mathrm{c.c.} \bigg>_{\Omega,Q}.
\label{chiD}
\end{align}
The diagram with a self-energy insertion
is presented in Fig.~\ref{fig3SM}E. In the $g_1^2$ channel, this diagram yields
\begin{align}
\chi^{(E)}_{1}(q,\omega) = 4U^{2}(2k_{F})\bigg<{\cal K}^{(E)}_{+}(Q,\Omega, q,\omega)\Pi_{-}(Q,\Omega)  \bigg>_{\Omega, Q} + \mathrm{c.c.}\;,
\end{align}
where a quartic combination of the Green's function is defined as
\beq
{\cal K}^{(E)}_{+}(Q,\Omega, q,\omega)= \sum_{\alpha=\pm} \bigg<G_{+}^{2}(p,\epsilon)G_{+}(p+Q,\epsilon+\Omega)G_{+}(p + \alpha q,\epsilon +\alpha \omega)\bigg>_{\epsilon,p}\eeq
 and
 $\Pi_{-}(Q,\Omega)$ is given by Eq.~(\ref{pipm}). A convolution of ${\cal K}^{(E)}_+$ and $\Pi_-$ can be re-written as
 \begin{align}
\bigg< {\cal K}^{(E)}_{+}(Q,\Omega, q,\omega)\Pi_{-}(Q,\Omega) \bigg>_{\Omega,Q}
&=\sum_{\alpha = \pm} \bigg< -\frac{\alpha i}{{\bar \omega}{\bar \Omega}}\frac{1}{2\pi}\bigg[ \frac{q}{{\bar \omega}} - \frac{Q -\alpha q}{{\bar \Omega} -\alpha {\bar \omega}} \bigg]\bigg[\frac{i}{2\pi}\frac{Q}{{\bar \Omega}^{*}}\bigg] \bigg>_{\Omega,Q}
\\
&
= \bigg< \frac{2}{(2\pi)^2}\frac{Q}{{\bar \omega}^2{\bar \Omega}^{*}} \bigg[\frac{Q}{{\bar \Omega}} - \frac{Q-q}{{\bar \Omega}-{\bar \omega}} \bigg]\bigg>_{\Omega,Q}.\label{kppim}
\end{align}
In the last transformation, we used the symmetry of the integrand under $Q \rightarrow - Q$ and $\Omega \rightarrow -\Omega$. We again select terms that result in logarithms of the external variables, i.e., the second term in the last line of Eq.~(\ref{kppim}), and obtain
\begin{align}
\chi^{(E)}_{1}(q,\omega) = - U^{2}(2k_{F})\frac{8}{(2\pi)^2} \bigg< \frac{1}{{\bar \omega}^2}\frac{Q(Q-q)}{{\bar \Omega}^{*}({\bar \Omega}-{\bar \omega})} +
\mathrm{c.c.}
\bigg>_{\Omega,Q}.
\label{chiE}
\end{align}
Equations (\ref{chiD}) and (\ref{chiE}) contain an integral
\bea
I(q,\omega)&=&\bigg<\frac{Q(Q-q)}{{\bar \Omega}^{*}({\bar \Omega}-{\bar \omega})} \bigg>_{\Omega,Q}
= i\int\frac{dQ}{2\pi}\frac{Q(Q-q)}{2iv_{F}Q - {\bar \omega}}\left[ \theta(Q) - \theta(-Q+q) \right]
\nn\\
&
=&\frac{1}{2v_{F}}\frac{1}{2\pi} \bigg[\bigg( \frac{{\bar \omega}}{2iv_{F}} \bigg)^2 - \frac{{\bar \omega}}{2iv_{F}} q \bigg]\ln\bigg[\frac{(v_{F}\Lambda)^2}{\vert{\bar \omega}\vert^2}  \bigg] = -\frac{1}{(2v_{F})^3}\frac{1}{2\pi} \vert{\bar \omega}\vert^2\ln\bigg[\frac{(v_{F}\Lambda)^2}{\vert{\bar \omega}\vert^2}  \bigg].\label{int}
\eea
With the help of Eq.~(\ref{int}), we obtain for the combined contributions of diagrams D and E:
\bea
\chi^{(D)}_{1}(q,\omega) + \chi^{(E)}_{1}(q,\omega) &=& U^{2}(2k_{F})\frac{8}{(2\pi)^2} \left\{ \bigg[\frac{1}{\vert{\bar \omega}\vert^2} - \frac{1}{{\bar \omega}^2} \bigg]I(q,\omega)
+ \mathrm{c.c.} \right\}\nn\\
& =& -\frac{1}{4}\chi_{0}g_{1}^2 \frac{(v_{F}q)^2}{\vert{\bar \omega}\vert^2}\ln\bigg[\frac{(v_{F}\Lambda)^2}{\vert{\bar \omega}\vert^2}\bigg].
\label{DE}
\eea

As in the static case (see Ref.~\onlinecite{CM_prb03}),  all second-order diagrams containing at least one forward-scattering process, i.e., diagrams of order $g_{2}^2$, $g_{1}g_{2}$ , and $g^2_4$  cancel out. The Aslamazov-Larkin diagrams (not shown) vanish by spin-rotational symmetry. Therefore, the final result for the second-order contribution to the spin susceptibility is given by Eq.~(\ref{DE}), which coincides with Eq.~(\ref{result2}) of MT. We stress that the final result is valid for any value of the ratio
$|\omega/v_Fq|$ but when both $\omega$ and $v_Fq$ are much smaller than the cutoff energy scale.

\section{Renormalization group equations for the interaction vertices}

\begin{figure}[t]
\includegraphics[width=0.9 \linewidth]{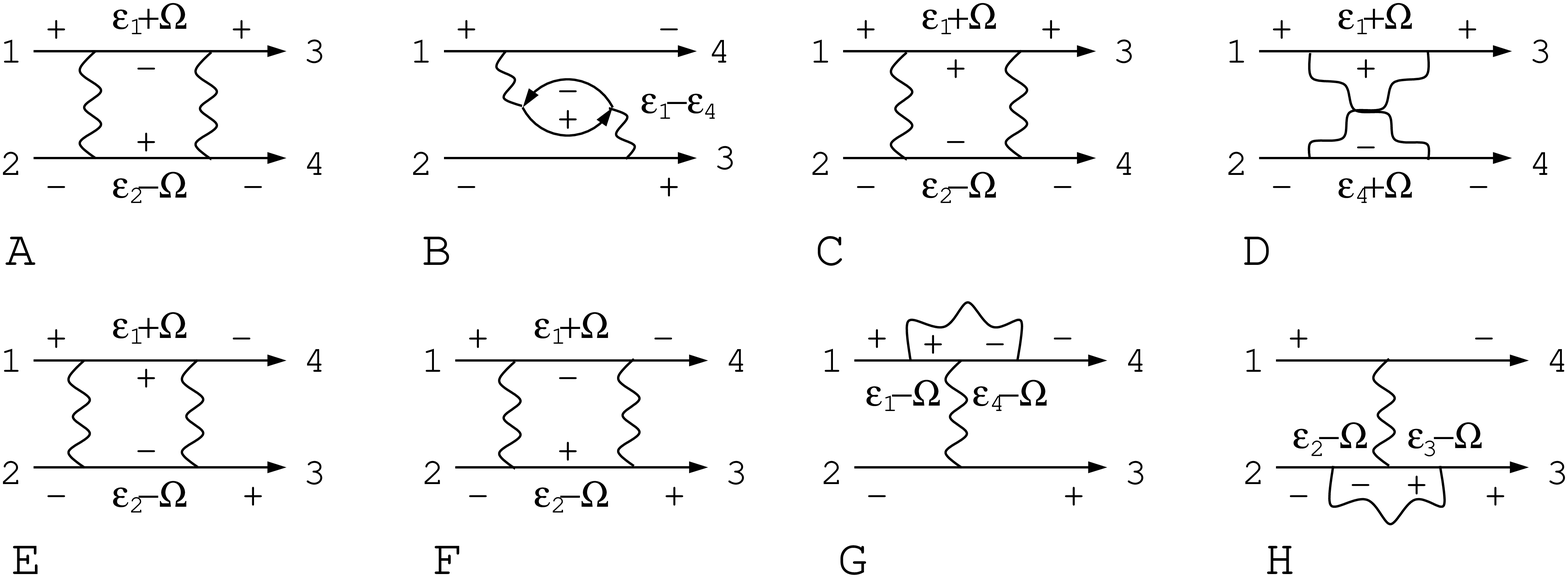}
\caption{
Diagrams for vertices
to second order in the interaction. Only
diagrams that result in logarithms are shown. Numerals $i$ with $i=1\dots 4$ on the incoming and outgoing legs correspond
to states with energies and momenta $(\varepsilon_i,p_i)$, such that $\varepsilon_1+\varepsilon_2=\varepsilon_3+\varepsilon_4$ and $p_1+p_2=p_3+p_4$.
 \label{fig5}}
\end{figure}

Renormalization group equations for the interaction vertices in 1D are usually derived for the case when the momenta of incoming
and outgoing fermions are on the Fermi surface (see, e.g. Refs.~\onlinecite{solyom,Giamarchi,M_review} of the MT).
To calculate the dynamic spin susceptibility, one needs to know the vertices away from the Fermi surface. In this section, we generalize the derivation for arbitrary relation between the momenta and frequencies of the incoming and outgoing fermions. As usual, we are interested only in those diagrams for the vertex that produce logarithms of external frequencies/momenta normalized by the cutoff.
These logarithms result from diagrams that contain either a Cooper or a $2k_F$ particle-hole bubble. All such diagrams are
shown in
Fig. ~\ref{fig5}.
Explicitly,
\begin{align}
\Gamma_{A} = U^{2}(2k_{F})\bigg<G_{-}(p_{1}+Q,\varepsilon_1+\Omega)G_{+}(p_{2}-Q,\varepsilon_2-\Omega) \bigg>_{\Omega,Q} = \frac{U^{2}(2k_{F})}{4\pi v_{F}}\ln\bigg[\frac{(v_{F}\Lambda)^2}{(\varepsilon_1+\varepsilon_2)^2 + v_{F}^2(p_{1}+p_{2})^2}\bigg],
\end{align}
\begin{align}
\Gamma_{B} = -2U^{2}(2k_{F}) \bigg< G_{+}(Q + p_{1}-p_{4},\Omega + \varepsilon_1 - \varepsilon_4)G_{-}(Q,\Omega) \bigg>_{\Omega,Q} = 2\frac{U^{2}(2k_{F})}{4\pi v_{F}}\ln\bigg[\frac{(v_{F}\Lambda)^2}{(\varepsilon_1-\varepsilon_4)^2 + v_{F}^2(p_{1}-p_{4})^2}\bigg],
\end{align}
\begin{align}
\Gamma_{C} = U^{2}(0) \bigg< G_{+}(p_{1}+Q,\varepsilon_1+\Omega)G_{-}(p_{2}-Q,\varepsilon_2-\Omega) \bigg>_{\Omega,Q} = \frac{U^2(0)}{4\pi v_{F}}\ln\bigg[\frac{(v_{F}\Lambda)^2}{(\varepsilon_1+\epsilon_2)^2 + v_{F}^2(p_{1}+p_{2})^2}\bigg],
\end{align}
\begin{align}
\Gamma_{D} = U^{2}(0) \bigg< G_{+}(p_{1}+Q,\varepsilon_1+\Omega)G_{-}(p_{4}+Q,\varepsilon_4+\Omega) \bigg>_{\Omega,Q} =
-\frac{U^2(0)}{4\pi v_{F}}\ln\bigg[\frac{(v_{F}\Lambda)^2}{(\varepsilon_1-\varepsilon_4)^2 + v_{F}^2(p_{1}-p_{4})^2}\bigg],
\end{align}
\begin{align}
\Gamma_{E}=\Gamma_{F} =U(0)U(2k_F)\bigg< G_+(p_1+Q,\varepsilon_1+\Omega)G_{-}(p_2-Q,\varepsilon_2-\Omega)\bigg>_{\Omega,Q}= \frac{U(0)U(2k_{F})}{4\pi v_{F}}\ln\bigg[\frac{(v_{F}\Lambda)^2}{(\varepsilon_1+\varepsilon_2)^2 + v_{F}^2(p_{1}+p_{2})^2}\bigg],
\end{align}
\begin{align}
\Gamma_{G} = U(0)U(2k_{F}) \bigg< G_{+}(p_{1}+Q,\varepsilon_1+\Omega)G_{-}(p_{4}+Q,\varepsilon_4+\Omega) \bigg>_{\Omega,Q} =
-\frac{U(0)U(2k_{F})}{4\pi v_{F}}\ln\bigg[\frac{(v_{F}\Lambda)^2}{(\varepsilon_1-\varepsilon_4)^2 + v_{F}^2(p_{1}-p_{4})^2}\bigg],
\end{align}
\begin{align}
\Gamma_{H} =U(0)U(2k_{F}) \bigg< G_{-}(p_{2}+Q,\varepsilon_2+\Omega)G_{+}(p_{3}+Q,\epsilon_3+\Omega) \bigg>_{\Omega,Q}
=-\frac{U(0)U(2k_{F})}{4\pi v_{F}}\ln\bigg[\frac{(v_{F}\Lambda)^2}{(\varepsilon_2-\varepsilon_3)^2 + v_{F}^2(p_{2}-p_{3})^2}\bigg].
\end{align}
A factor of $2$ in $\Gamma_{B}$ is due to a trace over spins.

We observe that all diagrams proportional to $U(0)U(2k_{F})$ add up to the following expression
\begin{align}
\Gamma_{E}+\Gamma_{F}+\Gamma_{G}+\Gamma_{H} = \frac{U(0)U(2k_{F})}{4\pi v_{F}}\ln\bigg[\frac{(\varepsilon_1-\varepsilon_4)^2 + v_{F}^2(p_{1}-p_{4})^2}{(\varepsilon_1+\varepsilon_2)^2 + v_{F}^2(p_{1}+p_{2})^2}\bigg],
\end{align}
which does not flow in the RG sense. The diagrams proportional to $U^2(0)$ are combined into
\begin{align}
\Gamma_{C}+\Gamma_{D} = \frac{U^2(0)}{4\pi v_{F}}\ln\bigg[\frac{(\varepsilon_1-\varepsilon_4)^2 + v_{F}^2(p_{1}-p_{4})^2}{(\varepsilon_1+\varepsilon_2)^2 + v_{F}^2(p_{1}+p_{2})^2}\bigg],
\end{align}
which does not flow in the RG sense as well. The remaining vertices, $\Gamma_{A}$ and $\Gamma_{B}$, are proportional
to $U(2k_F)^2 $, which, as is obvious from diagrams for $\Gamma_A$ and $\Gamma_B$ in Fig.~\ref{fig5}, must replaced by $g_1^2$.  It is also obvious that $\Gamma_{A}$ renormalizes
$g_2$ while $\Gamma_{B}$ renormalizes $g_{1}$. Denote
\beq
\ell \equiv \frac 12\ln\left[\frac{(v_{F}\Lambda)^2}{\vert {
\epsilon+iv_Fp
}\vert^2}\right]
\eeq
with
$\epsilon$ and $p$ being dummy variables. Then the vertices can be re-written as $\Gamma_{A} = (\pi v_{F})
{
g_1
^2}
{\bar \ell}/2$ and $\Gamma_{B} = (\pi v_{F})g_{1}^2 \ell$, where
one substitutes
$\epsilon=\varepsilon_1+\varepsilon_2$ and $p=p_1+p_2$ into $\ell$,  and $\epsilon=\varepsilon_1-\varepsilon_4$ and $p=p_1-p_4$ into ${\bar \ell}$.
Replacing the bare couplings by the running ones [$g_1\to \gamma_1(\ell)$ and $g_2\to \gamma_2(\bar\ell)$] in $\Gamma_{A}$ and $\Gamma_B$,  we arrive at the RG flow equations
\begin{align}
&\frac{d\gamma_{1}(\ell)}{d\ell} = - \gamma^2_{1}(\ell),\label{11}
\\
&\frac{d\gamma_{2}({\bar \ell})}{d{\bar \ell}} = - \frac{1}{2}\gamma^2_{1}({\bar \ell}).\label{21}
\end{align}
We see that $\gamma_1$ and $\gamma_2$ evolve in independent RG times.
Solving Eqs.~(\ref{11}) and (\ref{21}), we obtain
\begin{align}
\gamma_{1}(\ell) = \frac{g_{1}}{1+g_{1}\ell}, ~~\gamma_{1}({\bar \ell}) - 2\gamma_{2}({\bar \ell}) = \mathrm{const}\;.
\end{align}
In
what follows, we will need
$\gamma_{1}$, which is related to the spin part of the interaction vertex.
An explicit expression for $\gamma_1$ is
\begin{align}
\gamma_{1} = \frac{g_{1}}{1+\frac{g_{1}}{2}\ln\left[\frac{(v_{F}\Lambda)^2}{\left(\epsilon_1-\epsilon_4\right)^2 + v^2_{F}\left(p_1-p_4\right)^2}\right]}.
\end{align}

\section{
Spin susceptibility with renormalized interaction vertices}
In this section, we derive the final result for the spin susceptibility--Eq.~(\ref{result6}) of MT. Performing the perturbation theory in bare interaction up to second order, we showed that the logarithmic correction to spin susceptibility comes only from the backscattering processes. In the previous section, we derived
and solved the RG equation for the renormalized backscattering vertex, $\gamma_1$.
In this section, we calculate the skeleton diagrams in Fig.~\ref{fig4}B,C using $\gamma_1$ as the effective interaction.

We first consider the ladder-type diagram shown in Fig. \ref{fig4}C, which is  explicitly given by
\bea
\chi_{4\mathrm{C}}(q,\omega)
& =&
4\pi v_{F}
 \bigg< G_{+}(p,\epsilon)G_{+}(p+q,\epsilon+\omega)G_{-}(k,\Omega)G_{-}(k+q,\Omega+\omega)\left(
\gamma_{1}(p-k,\epsilon-\Omega)\right) \bigg> _{p,\epsilon,k,\Omega}
\nn\\
& =&
 \frac{
4\pi v_F}{\vert {\bar \omega} \vert^2}\bigg<\bigg[ G_{+}(p,\epsilon) - G_{+}(p+q,\epsilon+\omega)\bigg]\bigg[ G_{-}(k,\Omega) - G_{-}(k+q,\Omega+\omega)\bigg]
\gamma_{1}(p-k,\epsilon-\Omega)\bigg>_{p,\epsilon,k,\Omega}
\nn\\
& = &
 \frac{4\pi v_{F}}{\vert {\bar \omega} \vert^2}  \bigg<   \bigg[G_{+}(p,\epsilon)G_{-}(k+q,\Omega+\omega) + G_{+}(p+q,\epsilon+\omega)G_{-}(k,\Omega)
\nn\\
&&
-
G_{+}(p,\epsilon)G_{-}(k,\Omega)
- G_{+}(p+q,\epsilon+\omega)G_{-}(k+q,\Omega+\omega)\bigg] \gamma_{1}(p-k,\epsilon-\Omega)\bigg> _{p,\epsilon,k,\Omega},
\eea
where $\bar\omega$ is defined in Eq.~(\ref{baro}).
In the first line, we added up contributions resulting from two possible choices of assigning $\pm$ signs to the Green's functions to the left and to the right of the vertex. The product of the Green's functions in the first term in the last line does not depend on the external frequency and momenta. The second term in the last line is reduced to the first one by shifting the internal momenta and frequencies by the external ones:
$p+q\to p$, $k+q\to k$, etc.  After that, the dependence on the external variables remains only in the overall prefactor $1/|\bar\omega|^2$. Obviously, the last two terms cannot give a logarithmic dependence on the external variables and we omit them. Using that $\gamma_1(\l,\nu)$ is an even function of its both arguments, and re-labeling $k=p-Q$ and $\epsilon-\Omega\to \Omega$, the remaining terms  can be re-written as
\begin{align}
\chi_{4\mathrm{C}}(q,\omega)=
 4\frac{\pi v_{F}}{\vert {\bar \omega} \vert^2} \bigg< \bigg[G_{+}(p-q,\epsilon-\omega) + G_{+}(p+q,\epsilon+\omega)\bigg]G_{-}(p+Q,\Omega+\epsilon) \gamma_{1}(Q,\Omega)
\bigg> _{p,\epsilon,Q,\Omega}.
\label{C2}
\end{align}
We note, however, that in order to obtain a result valid to first order in $g_{1}$, one has to keep all the terms and recall that there is a cutoff in the problem.

For the sum of the self-energy
diagram
in Fig.~\ref{fig4}B and its mirror image
we obtain
\begin{align}
\chi_{4\mathrm{B}}(q,\omega)
&= 2\pi v_F
\sum_{\alpha=\pm} \bigg< G_{+}^{2}(p,\epsilon)G_{+}(p+
\alpha q,\epsilon
+\alpha \omega)G_{-}(k,\Omega) \gamma_{1}(p-k,\epsilon-\Omega) \bigg> _{p,\epsilon,k,\Omega} + \text{c.c.}\nn
\\
&
= \frac{2\pi v_{F}}{{\bar \omega}^2} \sum_\alpha \bigg<\bigg[ G_{+}(p,\epsilon) - G_{+}(p
 +\alpha q,\epsilon
 +\alpha  \omega)  \bigg]G_{-}(k,\Omega)\gamma_{1}(p-k,\epsilon-\Omega)\bigg> _{p,\epsilon,k,\Omega} + \text{c.c.}\nn
\\
&
=
-
 \frac{2  \pi v_{F}}{{\bar \omega}^2} \sum_\alpha\bigg< G_{+}(p
+\alpha q,\epsilon
+\alpha \omega)G_{-}(k,\Omega)\gamma_{1}(p-k,\epsilon-\Omega)\bigg> _{p,\epsilon,k,\Omega} + \text{c.c.}\nn\\
&=
 -\frac{2  \pi v_{F}}{{\bar \omega}^2} \sum_\alpha\bigg< G_{+}(p
+\alpha q,\epsilon
+\alpha \omega)G_{-}(p+Q,\Omega+\epsilon)\gamma_{1}(Q,\Omega)\bigg> _{p,\epsilon,Q,\Omega} + \text{c.c.},
\end{align}
where at the last step  we made the same change of variables as before Eq.~(\ref{C2}).
In going from the second line to the third, we discarded a term that does not
result in a logarithm of external variables.

Next, we need to add up the ladder and self-energy contributions and calculate the remaining integrals. Both contributions contain the same integral over
$\epsilon$ and
$p$, which is nothing else but the $2k_F$ polarization bubble with a characteristic log singularity:
\begin{align}
 \bigg< G_{+}(p \pm q,\epsilon \pm \omega)G_{-}(p+Q,\epsilon + \Omega) \bigg>_{\epsilon,p}
= -\frac{1}{4\pi v_{F}} \ln
\frac{(v_{F}\Lambda)^2}{\vert{\bar\Omega} \mp {\bar \omega} \vert^2},
\end{align}
where ${\bar\Omega}$ is defined in Eq.~(\ref{barO}).
The total logarithmic correction to the spin susceptibility is then given by
\begin{align}
\chi^{(\ln)}(q,\omega) =
\chi
_{4\text{B}}(q,\omega)+\chi_{4\text{C}}(q,\omega)
 &=  -
 \frac{1}{2} \bigg[\frac{2}{\vert {\bar \omega} \vert^2} - \frac{1}{{\bar \omega}^2} - \frac{1}{({\bar \omega}^{*})^2}   \bigg] \sum_\alpha   \bigg<\ln\bigg[\frac{(v_{F}\Lambda)^2}{\vert {
 \tilde \Omega}
 +\alpha {\bar \omega}\vert^2}\bigg]\gamma_{1}(k,\Omega) \bigg>_{k,\Omega}
\\
&= -2
\frac{
(v_{F}q)^2}{\vert{\bar\omega}\vert^4}
\sum_\alpha\bigg<\ln\bigg[\frac{(v_{F}\Lambda)^2}{\vert {\bar \Omega}
+\alpha {\bar \omega}\vert^2}\bigg]\gamma_{1}(Q,\Omega) \bigg>_{Q,\Omega} .
\end{align}
Remaining
integration over $\Omega$ and $Q$
can be carried out by
by switching to polar coordinates, such that $\Omega = r\sin\phi$ and $v_{F}Q= r\cos\phi$. Defining
${\vec R} \equiv (v_{F}q, \omega)$ and $\phi_{R} \equiv \arctan(\omega/v_{F}q)$, we can then re-write the argument of the logarithm as
\begin{align}
\vert {\bar \Omega} \pm {\bar \omega} \vert^2 = (\Omega \pm \omega)^2 + v_{F}^2(Q \pm q)^2 = ({\vec r} \pm {\vec R})^2 = r^2 + R^2 \pm 2rR\cos(\phi - \phi_{R}).
\end{align}
The angle $\phi_R$ can be eliminated by a shift $\phi-\phi_R\to \phi$. Summation over $\alpha=\pm$ in the expression for the spin susceptibility results in a factor of two, with that we reproduce the first line in Eq.~(\ref{result6}) of MT. In new variables, the renormalized interaction $\gamma_{1}$, derived in the previous section, reads
\begin{align}
\gamma_{1}(r)  = \frac{g_{1}}{1+\frac{g_{1}}{2}\ln
\frac{(v_{F}\Lambda)^2}{r^2}
}.
\end{align}
After transformation to polar coordinates, we impose a cutoff on the radial variable $r$: $0\leq r \leq v_F\Lambda'$.
The two cutoffs--$\Lambda$ and $\Lambda'$--do not have to coincide but, to leading logarithmic approximation, their ratio does not enter the final result and in MT we chose $\Lambda'=\Lambda$.
On changing the variables and summing over $\alpha$, we obtain
\begin{align}
\chi^{(\ln)}(q,\omega) &= -
\chi_0
\frac{(v_{F}q)^2}{\vert{\bar\omega}\vert^4}
\int_{0}^{v_{F}\Lambda'} 
{drr} \gamma_{1}(r)J(r,R),
\label{chiln1}
\end{align}
where
\beq
J(r,R)=\int_{0}^{2\pi} \frac{d\phi}{2\pi}\ln\bigg[\frac{(v_{F}\Lambda)^2}{r^2+R^2 + 2rR\cos\phi}\bigg].
\eeq
We will need the following integral:
\bea
&&\int_{0}^{2\pi} \frac{d\phi}{2\pi}\ln\left(A + \cos \phi\right) = \int^A_1 dA' \int_{0}^{2\pi}\frac{d\phi}{2\pi} \frac{1}{A' + \cos\phi}-\int^{2\pi}_0d\phi \ln\left(1+\cos\phi\right)\nn\\
& =& \int^A_1 dA' \frac{
1}{\sqrt{(A')^2 - 1}}-\ln 2 = \ln\frac{A+\sqrt{A^2 - 1}}{2}
\label{intA}
\eea
for $A>1$, which in our case is always satisfied.
With the help of Eq.~(\ref{intA}), the radial integral in Eq.~(\ref{chiln1}) becomes
\begin{align}
&\int_{0}^{v_{F}\Lambda'} rdr~\ln\bigg[\frac{2(v_{F}\Lambda)^2}{r^2+R^2+\vert r^2 - R^2 \vert}\bigg]\gamma_{1}(r)\nn
\\
&= \frac{1}{2}
\left\{\ln\bigg[\frac{(v_{F}\Lambda)^2}{
R^2}\bigg] \int_{0}^{R^2}dx \frac{g_{1}}{1+\frac{g_{1} }{2}\ln\left[\frac{(v_{F}\Lambda)^2}{x} \right]  }  + \int_{R^2}^{(v_{F}\Lambda)^2} dx \ln\bigg[\frac{(v_{F}\Lambda)^2}{
x}\bigg]\frac{g_{1}}{1+\frac{g_{1} }{2}\ln\left[\frac{(v_{F}\Lambda)^2}{x} \right]  }
\right\}.
\label{chiln2}
\end{align}
The integrals in the last equation can be expressed via the integral exponential function [Ei$(x)$]. However, as we only need the final result to leading logarithmic
 approximation, it is more convenient to expand the fractions in Eq.~(\ref{chiln2}) into geometric series, solve the integrals order by order, select the
  leading logarithmic terms, and sum the series back. To order $g_1$, the$R^2\ln R^2$ terms from the two integrals cancel each other, and the leading contribution is of order $R^2g_1$; to order $g_1^2$,  the $R^2\ln^2 R^2$ terms cancel out but the $R^2\ln R^2$ terms survive, etc. One also need to discard $R$-independent ultraviolet terms,
  proportional to $\Lambda^2$, as they must cancel out with other ultraviolet contributions we discarded at previous steps. The resulting series give
\begin{align}
&
-R^2g_{1}  + R^2\frac{g_{1}^2}{2}\ln
\frac{(v_{F}\Lambda)^2}{R^2}
 - R^2\frac{g_{1}^{3}}{4}\ln^2
 \frac{(v_{F}\Lambda)^2}{R^2}
 + ...
\nonumber\\
&\rightarrow -
\frac{R^2 g_{1}}{1+\frac{g_{1}}{2}\ln
\frac{(v_{F}\Lambda)^2}{R^2}
},
\end{align}
Recalling now that $R^2 = \vert{\bar \omega} \vert^2 = \omega^2 + (v_{F}q)^2$,
we obtain the final result for logarithmic correction to
the spin susceptibility
\begin{align}
\chi^{(\ln)}(q,\omega) &
= \frac{1}{2}\chi_{0}
 \frac{(v_{F}q)^2}{\omega^2+(v_Fq)^2
 } \frac{
g_{1}}{1+\frac{g_{1}}{2}\ln
\frac{(v_{F}\Lambda)^2}{\vert{\bar\omega}\vert^2}
},
\end{align}
which
coincides with Eq.~(\ref{result6}) of MT.
which
coincides with Eq.~(\ref{result6}) of MT.
In order to recover full
Eq.~(\ref{result5}) with non-logarithmic terms proportional to $g_{1}$, one has to keep
those combinations of Green functions
that
do not contribute to the geometric progression leading to the result above, i.e.,
those combinations
we have omitted in the beginning of this Appendix. It is important to remember that,
since the model has finite cutoff,
every shift of the momentum under the integral also shifts the cutoff.

\section{Bosonization procedure}
In this section, we review some details of the bosonization procedure with the aim of deriving a non-local form of the sine-Gordon Hamiltonian
in Eq.~( \ref{backscattering}) of MT and also clarifying an ambiguity regarding the parameters of the Gaussian part (see comment \onlinecite{comment_giamarchi} of MT.)

As usual, the fermion operators are represented as \
\begin{align}
\Psi_{\alpha}(x) = e^{ik_{F}x}R_{\alpha}(x) + e^{-ik_{F}x}L_{\alpha}(x),
\end{align}
where $R_{\alpha}$ and $L_{\alpha}$ are spin $\alpha=\uparrow,\downarrow$ fermion operators describing right and left movers, correspondingly.
At the next step, one linearizes the spectrum close to Fermi energy and then writes Hamiltonian of non-interacting fermions as
\begin{align}
H_{0} = -\frac{i}{2\pi}v_{F} \sum_{\alpha = \uparrow,\downarrow}\int dx \left[ :R_{\alpha}^{\dag}(x)\partial_{x}R_{\alpha}(x):  - :L_{\alpha}^{\dag}(x)\partial_{x}L_{\alpha}(x):\right].
\end{align}
For further derivations, let us introduce the current operators as
\begin{align}
&J^{(c)}_{R} = \frac{1}{2}\left(R_{\uparrow}^{\dag}R_{\uparrow} + R_{\downarrow}^{\dag}R_{\downarrow} \right), ~~ J^{(c)}_{L} = \frac{1}{2}\left(L_{\uparrow}^{\dag}L_{\uparrow} + L_{\downarrow}^{\dag}L_{\downarrow} \right)
\nonumber\\
&J^{(z)}_{R} = \frac{1}{2}\left(R_{\uparrow}^{\dag}R_{\uparrow} - R_{\downarrow}^{\dag}R_{\downarrow} \right), ~~ J^{(z)}_{L} = \frac{1}{2}\left(L_{\uparrow}^{\dag}L_{\uparrow} - L_{\downarrow}^{\dag}L_{\downarrow} \right)
\nonumber\\
&J^{(x)}_{R} = \frac{1}{2}\left(R_{\uparrow}^{\dag}R_{\downarrow} + R_{\downarrow}^{\dag}R_{\uparrow} \right),~~
J^{(x)}_{L} = \frac{1}{2}\left(L_{\uparrow}^{\dag}L_{\downarrow} + L_{\downarrow}^{\dag}L_{\uparrow} \right)
\nonumber\\
&J^{(y)}_{R} = \frac{i}{2}\left(R_{\uparrow}^{\dag}R_{\downarrow} - R_{\downarrow}^{\dag}R_{\uparrow} \right),~~
J^{(y)}_{L} = \frac{i}{2}\left(L_{\uparrow}^{\dag}L_{\downarrow} - L_{\downarrow}^{\dag}L_{\uparrow} \right),
\end{align}
in the following we will be referring to ${\vec J}_{R/L}$ as a spin current vector with $(J^{x}_{R/L},J^{y}_{R/L},J^{z}_{R/L})$ components. The interaction part of the Hamiltonian is
\begin{align}
H_{\text{int}} = \frac{1}{2}\int dx dx^\prime V(x-x^\prime)\rho(x)\rho(x^\prime),
\end{align}
where $\rho(x) =\sum_{\alpha}\rho_{\alpha}(x) = \sum_{\alpha}\Psi_{\alpha}^{\dag}(x)\Psi_{\alpha}(x)$ is fermion density.
We introduce density for a given spin as a sum of $0$ and $2k_{F}$ parts as $\rho_{\alpha}(x) = \rho_{\alpha}^{(0)}(x) + \rho_{\alpha}^{(2k_{F})}(x)$. For the forward-scattering part of $H_{\text{int}}$, we then obtain
\begin{align}
&\rho_{\uparrow}^{(0)}(x)\rho_{\uparrow}^{(0)}(x^{\prime}) + \rho_{\downarrow}^{(0)}(x)\rho_{\downarrow}^{(0)}(x^{\prime}) =
\nonumber\\
&= 2\left(  J^{(c)}_{R}(x)J^{(c)}_{R}(x^{\prime}) + J^{(c)}_{L}(x)J^{(c)}_{L}(x^{\prime}) + J^{(c)}_{R}(x)J^{(c)}_{L}(x^{\prime}) + J^{(c)}_{L}(x)J^{(c)}_{R}(x^{\prime})  \right)
\nonumber\\
&+ 2\left(  J^{(z)}_{R}(x)J^{(z)}_{R}(x^{\prime}) + J^{(z)}_{L}(x)J^{(z)}_{L}(x^{\prime}) + J^{(z)}_{R}(x)J^{(z)}_{L}(x^{\prime}) + J^{(z)}_{L}(x)J^{(z)}_{R}(x^{\prime})  \right),
\end{align}
and
\begin{align}
&\rho_{\downarrow}^{(0)}(x)\rho_{\uparrow}^{(0)}(x^{\prime}) + \rho_{\uparrow}^{(0)}(x)\rho_{\downarrow}^{(0)}(x^{\prime}) =
\nonumber\\
&= 2J_{R}^{(c)}(x)J_{R}^{(c)}(x^{\prime}) - 2J_{R}^{(z)}(x)J_{R}^{(z)}(x^{\prime})
+2J_{L}^{(c)}(x)J_{R}^{(c)}(x^{\prime}) - 2J_{L}^{(z)}(x)J_{R}^{(z)}(x^{\prime})
\nonumber\\
&
+2J_{R}^{(c)}(x)J_{L}^{(c)}(x^{\prime}) - 2J_{R}^{(z)}(x)J_{R}^{(z)}(x^{\prime})
+2J_{L}^{(c)}(x)J_{L}^{(c)}(x^{\prime}) - 2J_{L}^{(z)}(x)J_{L}^{(z)}(x^{\prime}).
\end{align}
Adding up the two expressions above yields
\begin{align}
&\rho_{\uparrow}^{(0)}(x)\rho_{\uparrow}^{(0)}(x^{\prime}) + \rho_{\downarrow}^{(0)}(x)\rho_{\downarrow}^{(0)}(x^{\prime})+\rho_{\downarrow}^{(0)}(x)\rho_{\uparrow}^{(0)}(x^{\prime}) + \rho_{\uparrow}^{(0)}(x)\rho_{\downarrow}^{(0)}(x^{\prime})
\nonumber\\
&
= 4\left[ J_{R}^{(c)}(x)J_{R}^{(c)}(x^{\prime}) + J_{L}^{(c)}(x)J_{R}^{(c)}(x^{\prime}) + J_{R}^{(c)}(x)J_{L}^{(c)}(x^{\prime}) + J_{L}^{(c)}(x)J_{L}^{(c)}(x^{\prime})\right],
\end{align}
from which we see that only the charge sector is affected by forward scattering.
A product of two $2k_{F}$-components of the densities becomes
\begin{align}
&\rho_{\uparrow}^{(2k_{F})}(x)\rho_{\uparrow}^{(2k_{F})}(x^{\prime}) + \rho_{\downarrow}^{(2k_{F})}(x)\rho_{\downarrow}^{(2k_{F})}(x^{\prime}) =
\nonumber\\
& =-\Psi_{R\uparrow}^{\dag}(x)\Psi_{R\uparrow}(x^{\prime})
\Psi_{L\uparrow}^{\dag}(x^{\prime})\Psi_{L\uparrow}(x)
-\Psi_{L\uparrow}^{\dag}(x)\Psi_{L\uparrow}(x^{\prime})
\Psi_{R\uparrow}^{\dag}(x^{\prime})\Psi_{R\uparrow}(x)
\nonumber\\
&-\Psi_{R\downarrow}^{\dag}(x)\Psi_{R\downarrow}(x^{\prime})
\Psi_{L\downarrow}^{\dag}(x^{\prime})\Psi_{L\downarrow}(x)
-\Psi_{L\downarrow}^{\dag}(x)\Psi_{L\downarrow}(x^{\prime})
\Psi_{R\downarrow}^{\dag}(x^{\prime})\Psi_{R\downarrow}(x)
\nonumber\\
&
+(4k_{F}),
\end{align}
where $(4k_F)$ stands for $4k_F$ scattering terms which vanish in the absence of lattice.
The remaining backscattering part of the interaction is given by
\begin{align}
&\rho_{\downarrow}^{(2k_{F})}(x) \rho_{\uparrow}^{(2k_{F})}(x^{\prime}) + \rho_{\uparrow}^{(2k_{F})}(x) \rho_{\downarrow}^{(2k_{F})}(x^{\prime})
\nonumber\\
&
=\Psi_{R\downarrow}^{\dag}(x)\Psi_{L\downarrow}(x)\Psi_{L\uparrow}^{\dag}(x^{\prime})\Psi_{R\uparrow}(x^{\prime})
+ \Psi_{L\downarrow}^{\dag}(x)\Psi_{R\downarrow}(x)\Psi_{R\uparrow}^{\dag}(x^{\prime})\Psi_{L\uparrow}(x^{\prime})
\nonumber\\
&
+\Psi_{R\uparrow}^{\dag}(x)\Psi_{L\uparrow}(x)\Psi_{L\downarrow}^{\dag}(x^{\prime})\Psi_{R\downarrow}(x^{\prime})
+ \Psi_{L\uparrow}^{\dag}(x)\Psi_{R\uparrow}(x)\Psi_{R\downarrow}^{\dag}(x^{\prime})\Psi_{L\downarrow}(x^{\prime})
\nonumber\\
&
+(4k_{F}).
\end{align}

In the limit when $U(x)$ is local,
one can show that the Hamiltonian
of the spin sector assumes a Sugawara form \cite{GNT}
\begin{equation}
H = \frac{2\pi v_{F}}{3} \int dx~ \left[ :{\vec J}_{R}{\vec J}_{R}: + :{\vec J}_{L}{\vec J}_{L}: \right] + g_{\text{bs}}\int dx :{\vec J}_{R}{\vec J}_{L}:
\end{equation}
We can see that in this description the Hamiltonian is manifestly
$SU(2)$ invariant. Also, $v_{F}$ does not get renormalized by interactions in this case.

In another approach, non-Abelian bosonization, one writes spin $\alpha$ Fermion operators in a coherent state representation as
\begin{align}
& R_{\alpha}(x) = \frac{\eta_{\alpha}\sqrt{\Lambda_{F}}}{\sqrt{2\pi}} e^{i\phi_{+,\alpha}(x)} ,
\nonumber\\
& L_{\alpha}(x) = \frac{\eta_{\alpha}\sqrt{\Lambda_{F}}}{\sqrt{2\pi}} e^{i\phi_{-,\alpha}(x)} ,
\end{align}
where $\eta_{\alpha}$ is a Klein factor ensuring fermion nature of the field operator, as usual $\phi_{\pm,\alpha}(x) = \mp \phi_{\alpha}(x) + \theta_{\alpha}(x)$, where $\phi(x)$ and $\theta(x)$ are canonically conjugate boson operators, such as $[\phi(x),\partial_{y}\theta(y)] = i\pi\delta(x-y)$. For a review see, for example, Ref. [\onlinecite{Giamarchi}] of the MT.

The densities of the left/right movers are then introduced as
\begin{align}
&\rho_{\pm,\alpha} = \frac{1}{2\pi} \bigg( \nabla\phi_{\alpha} \mp \nabla\theta_{\alpha} \bigg),
\nonumber\\
&
\rho_{\alpha} = \rho_{+,\alpha} + \rho_{-,\alpha} = \frac{1}{\pi} \nabla\phi_{\alpha},
\nonumber\\
&
\rho_{c} = \rho_{\uparrow} + \rho_{\downarrow} = \frac{\sqrt{2}}{\pi} \nabla\phi_{c},
\nonumber\\
&
\rho_{s} = \rho_{\uparrow} - \rho_{\downarrow} = \frac{\sqrt{2}}{\pi} \nabla\phi_{s},
\nonumber\\
&
\phi_{c/s} = \frac{\phi_{\uparrow} \pm \phi_{\downarrow}}{\sqrt{2}}, ~~ \theta_{c/s} = \frac{\theta_{\uparrow} \pm \theta_{\downarrow}}{\sqrt{2}}.
\end{align}
In terms of these densities, the free part of the
Hamiltonian
becomes
\begin{align}
H_{0} = \frac{v_{F}}{2\pi} \sum_{i=c,s}\int dx~ \bigg[ \left(\nabla_{x}\phi_{i}(x) \right)^2 +  \left(\nabla_{x}\theta_{i}(x) \right)^2 \bigg].
\end{align}
In the interaction part, one needs to be careful with treating the backscattering of fermions with the same spin.
This issue was addressed in the spineless case\cite{capponi,maslov} but not for fermions with spins.

In the forward-scattering part, one can safely take the limit of a local interaction from the very beginning:
\begin{align}
&\frac{1}{2}\int dxdx^\prime~U(x-x^\prime) \bigg[  \rho_{\uparrow}^{(0)}(x)\rho_{\uparrow}^{(0)}(x^{\prime}) + \rho_{\downarrow}^{(0)}(x)\rho_{\downarrow}^{(0)}(x^{\prime})+\rho_{\downarrow}^{(0)}(x)\rho_{\uparrow}^{(0)}(x^{\prime}) + \rho_{\uparrow}^{(0)}(x)\rho_{\downarrow}^{(0)}(x^{\prime}) \bigg]
\nn\\
&
= 2\int dxdx^\prime U(x-x^\prime)\bigg[ J_{R}^{(c)}(x)J_{R}^{(c)}(x^{\prime}) +  J_{L}^{(c)}(x)J_{L}^{(c)}(x^{\prime}) \bigg]
+  2\int dxdx^\prime U(x-x^\prime)\bigg[ J_{L}^{(c)}(x)J_{R}^{(c)}(x^{\prime})
+ J_{R}^{(c)}(x)J_{L}^{(c)}(x^{\prime}) \bigg]
\nn\\
&
\rightarrow \frac{v_{F} g_{4}}{2\pi}\int dX~ \left\{ \left[\nabla_{X} \phi_{c}(X)\right]^2 + \left[\nabla_X \theta_{c}(X)\right]^2 \right\}
+ \frac{v_{F} g_{2}}{2\pi}\int dX~ \left\{ \left[\nabla_{X} \phi_{c}(X)\right]^2 - \left[\nabla_X \theta_{c}(X)\right]^2 \right\},
\label{G1}
\end{align}
where we introduced new coordinates $X = (x + x^{\prime})/2$ and ${\text x} = x - x^\prime$
and made approximations: $\phi_{c}(x)\approx\phi_{c}(X)$ and $\phi_{c}(x^\prime)\approx\phi_{c}(X)$.
In the backscattering part, we treat the interaction as non-local and obtain
\begin{align}
& \frac{1}{2}\int dxdx^\prime~U(x-x^\prime)\left[ \rho_{\uparrow}^{(2k_{F})}(x)\rho_{\uparrow}^{(2k_{F})}(x^{\prime}) + \rho_{\downarrow}^{(2k_{F})}(x)\rho_{\downarrow}^{(2k_{F})}(x^{\prime}) +\rho_{\uparrow}^{(2k_{F})}(x)\rho_{\downarrow}^{(2k_{F})}(x^{\prime}) + \rho_{\downarrow}^{(2k_{F})}(x)\rho_{\uparrow}^{(2k_{F})}(x^{\prime})\right]
\nn\\
&
=\frac{\Lambda_F^2}{2\pi^2} \int \int dxdx^\prime ~ U(x-x^{\prime}) \cos\left\{\sqrt{2}\left[\phi_{s}(x) \pm \phi_{s}(x^{\prime})\right]\right\}
 \cos\left\{\sqrt{2}\left[\phi_{c}(x)-\phi_{c}(x^{\prime})\right] +2k_{F}(x-x^{\prime})\right\}.
\end{align}
Now we are going to take the local limit, treating
the cosine of the difference
$\phi_{s}(x) - \phi_{s}(x^{\prime})$
 in the following manner:
\begin{align}
&\cos\left\{\sqrt{2}\left[\phi_{s}(x) - \phi_{s}(x^{\prime})\right]\right\}
= :\cos\left\{\sqrt{2}\left[\phi_{s}(x) - \phi_{s}(x^{\prime})\right]\right\}:
\exp\left\{- \left< (\phi_{s}(x) - \phi_{s}(x^{\prime}) )^2 \right> \right\}
\nn\\
&
\approx \cos\left\{\sqrt{2}{\text x}\nabla_{X}\phi_{s}(X)\right\}\frac{1}{\Lambda_{F}\sqrt{{\text x}^2+ (v_{F}\tau)^2}} 
\approx \left\{ 1- {\text x}^2 \left[\nabla_{X} \phi_{s}(X)\right]^2 \right\} \frac{1}{\Lambda_{F}\sqrt{{\text x}^2+ (v_{F}\tau)^2}},
\end{align}
where correlation function $\left< \left[\phi_{s}(x) - \phi_{s}(x^{\prime}) \right]^2 \right> = - \frac{1}{2}\ln\left[ \frac{\Lambda_{F}^{-2}}{{\text x}^2 + (v_{F}\tau)^2} \right]$, and the same for correlations in charge sector.
Same procedure is applied to the charge part
\begin{align}
&\cos\left\{\sqrt{2}\left[\phi_{c}(x) - \phi_{c}(x^{\prime})\right] + 2k_{F}(x-x^\prime) \right\}
= :\cos\left\{\sqrt{2}\left[\phi_{c}(x) - \phi_{c}(x^{\prime})\right] + 2k_{F}(x-x^\prime)\right\}:
\exp\left\{- \left< (\phi_{c}(x) - \phi_{c}(x^{\prime}) )^2 \right> \right\}
\nn\\
&
\approx \cos\left\{\sqrt{2}{\text x}\nabla_{X}\phi_{c}(X) + 2k_{F}{\text x} \right\}\frac{1}{\Lambda_{F}\sqrt{{\text x}^2+ (v_{F}\tau)^2}}
\approx \left\{ 1- {\text x}^2
\left[\nabla_{X} \phi_{c}(X)\right]^2 \right\} e^{i2k_{F}{\text x}}\frac{1}{\Lambda_{F}\sqrt{{\text x}^2+ (v_{F}\tau)^2}},
\end{align}
where in the last line we used symmetry under ${\text x} \rightarrow -{\text x}$ and $X \rightarrow -X$.
The product of the gradient terms gives results in renormalization of the Gaussian part of the Hamiltonian:
\begin{align}
&\int \int d{\text x}dX ~ U\left({\text x}\right) e^{i2k_{F}{\text x}}  \frac{\Lambda_{F}^2}{2\pi^2} \left\{ 1- {\text x}^2 \left[\nabla_{X} \phi_{c}(X)\right]^2 \right\} \left\{ 1- {\text x}^2 \left[\nabla_{X} \phi_{s}(X)\right]^2 \right\} \frac{1}{\Lambda_{F}^2 \left[{\text x}^2+ (v_{F}\tau)^2\right]}
\nn\\
&
\approx -\frac{v_{F} g_{1\parallel}}{2\pi} \int dX \left\{ \left[\nabla_{X} \phi_{c}(X)\right]^2 + \left[\nabla_{X} \phi_{s}(X)\right]^2 \right\},
\label{G2}
\end{align}
where we kept only lowest gradient terms
and
defined $g_{1||}/\pi v_{F} = \int d{\text x} U({\text x})e^{i2k_{F}{\text x}}$.
The cosine of the sum $\phi_{s}(x) + \phi_{s}(x^{\prime})$ in the local limit
gives the familiar
sine-Gordon term
\begin{align}
\frac{2\Lambda_{F}^2 v_{F} g_{1\perp}}{4\pi} \int dX~\cos\left[2\sqrt{2}\phi_{s}(X)\right].
\end{align}

Combining the Gaussian parts from Eqs.~(\ref{G1}) and (\ref{G2}), we obtain the total Hamiltonian as a sum of the charge and spin parts:
Overall, the spin part of the Hamiltonian becomes
\bea
H&=&H_c+H_s,\nn\\
H_{c} &=& \frac{v_{c}}{2\pi}\int dx~ \bigg[ \frac{1}{K_{c}}\left(\nabla_{x}\phi_{c}(x) \right)^2 +  K_{c}\left(\nabla_{x}\theta_{c}(x) \right)^2 \bigg],\nn\\
H_{s} &=& \frac{v_{s}}{2\pi}\int dx ~\bigg[ \frac{1}{K_{s}}\left(\nabla_{x}\phi_{s}(x) \right)^2 +  K_{s}\left(\nabla_{x}\theta_{s}(x) \right)^2 \bigg] + \frac{2\Lambda_{F}^2 v_{F} g_{1\perp}}{4\pi} \int dx~\cos\left[2\sqrt{2}\phi_{s}(x)\right]
\eea
with
\bea
K_{c}& =& \sqrt{\frac{1+g_{4}-g_{2}}{1+g_{4}+g_{2}-g_{1\parallel}}},\nn\\
v_{c} &= &v_{F}\sqrt{(1+g_{4}-g_{2})(1+g_{4}+g_{2}-g_{1\parallel})},\nn\\
K_{s} &=& \frac{1}{\sqrt{1-g_{1\parallel}}}\nn\\
v_s&=&v_{F}\sqrt{1-g_{1\parallel}}.
\eea
As mentioned in the main text, the above expressions for the charge and spin velocities ensure
that the results for specific heat\cite{saha} and spin susceptibility of 1D system, obtained via bosonization
coincide with the perturbation theory in the fermion language.
\end{widetext}

\end{document}